\documentclass[fleqn,12pt]{article}
\usepackage{espcrc1}

\usepackage{psfig}
\usepackage{graphicx}
\usepackage[figuresright]{rotating}

\newcommand{\AmS}{{\protect\the\textfont2
  A\kern-.1667em\lower.5ex\hbox{M}\kern-.125emS}}

\title{Towards Modelling The Internet Topology -- The Interactive Growth Model
\thanks{This research is supported by the U.K.
        Engineering and Physical Sciences Research Council
         (EPSRC) under grant no. GR--R30136--01.}}

\author{Shi Zhou\address{Department of Electronic Engineering,
        Queen Mary, University of London\\
        Mile End Road, London, E1 4NS, United Kingdom}
         and Ra\'ul J. Mondrag\'on\addressmark}

\begin{document}

\maketitle

\begin{abstract}
The Internet topology at the Autonomous Systems level (AS graph)
has a power--law degree distribution and a tier structure. In this
paper, we introduce the Interactive Growth (IG) model based on the
joint growth of new nodes and new links. This simple and dynamic
model compares favorable with other Internet power--law topology
generators because it not only closely resembles the degree
distribution of the AS graph, but also accurately matches the
hierarchical structure, which is measured by the recently reported
rich-club phenomenon.
\end{abstract}

\section{INTRODUCTION}

Faloutsos~{\sl et~al} \cite{Falou99} discovered that the Internet
topology at the Autonomous Systems (ASes) level (AS graph) has a
power--law degree distribution, $P(k)\propto k^{-2.22}$, where
node degree $k$ is the number of links a node has.
Subramanian~{\sl et~al} \cite{Subra02},using a heuristic argument
based on the commercial relationship between ASes, found that the
Internet has a tier structure. Tier 1 consists of a `core' of ASes
which are well connected to each other. Recently the rich--club
phenomenon \cite{Zhou03} was introduced as a quantitative metric
to characterize the core tier without making any heuristic
assumption on the interaction between network elements.

There are a number of Internet power--law topology generators,
some are degree-based and others are structure-based.
Tangmunarunkit~{\sl et al} \cite{Tangm02} found that degree
distributions produced by structure-based generators are not
power--laws.

In this paper we introduce the Interactive Growth (IG) model based
on the joint growth of new nodes and new links. We compare the AS
graph against the IG model and three degree-based models, which
are the Barab\'asi and Albert (BA) scale--free model
\cite{Barab99}, the Inet--3.0 model \cite{Winick02} and the
Generalized Linear Preference (GLP) model \cite{Bu02}. We show
that the IG model compares favorable with other Internet
power--law topology generators because it not only closely
resembles the degree distribution of the AS graph, but also
accurately matches the hierarchical structure measured by the
rich-club phenomenon. The IG model is simple and dynamic, we
believe it is a good step towards modelling the Internet topology.

\section{RICH--CLUB PHENOMENON}

Power--law topologies have a small number of nodes having large
numbers of links. We call these nodes `rich nodes'. The AS graph
shows a rich--club phenomenon \cite{Zhou03}, in which rich nodes
are very well connected to each other and rich nodes are connected
preferentially to the other rich nodes.

Recently the rich--club phenomenon was measured \cite{Zhou03} in
the so--called original--maps and extended--maps of the AS graph.
The original--maps are based on the BGP routing tables collected
by the University of Oregon Route Views Project \cite{Oregon}. The
extended--maps \cite{Chen02} use additional data sources, such as
the Looking Glass (LG) data and the Internet Routing Registry
(IRR) data. The two maps have similar numbers of nodes, while the
extended--maps have $40\%$ more links than the original--maps. The
research showed the majority of the missing links in the
original--maps are connecting links \emph{between} rich nodes of
the extended--maps, and therefore, the extended--maps show the
rich--club phenomenon significantly stronger than the
original--maps.

The rich-club phenomenon is relevant because the connectivity
between rich nodes can be crucial for network properties, such as
network routing efficiency, redundancy and robustness. For example
in the AS graph, the members of the rich-club are very well
connected to each other. This means that there are a large number
of alternative routing paths between the club members where the
average path length inside the club is very small (1 to 2 hops).
Hence, the rich-club acts as a super traffic hub and provides a
large selection of shortcuts. Network models without the rich-club
phenomenon may under--estimate the efficiency and flexibility of
the traffic routing in the AS graph. On the other hand, models
without the rich-club phenomenon may over--estimate the robustness
of the network to a node attack \cite{Albert00} where the removal
of a few of its richest club members can break down the network
integrity.

The rich--club phenomenon is a quantitatively simple way to
differentiate the tier structures between power--law topologies
and it provides a criterion for new network models.

\section{DEGREE--BASED INTERNET TOPOLOGY GENERATORS}

\subsection{Inet--3.0 model}

The Inet--3.0 model \cite{Winick02} was designed to match the
measurements of the original--maps of the AS graph. The model is
capable of creating networks with degree distribution similar to
that of the measurements. The number of links generated by the
model depends on two parameters, which are the total number of
nodes and the percentage of nodes with degree one. The model
typically generates 26\% less links than the extended--AS graph.

\subsection{Barab\'asi--Albert (BA) model}

The BA model \cite{Barab99} shows that a power--law degree
distribution can arise from two generic mechanisms: 1)
\emph{growth}, where networks expand continuously by the addition
of new nodes, and 2) \emph{preferential attachment}, where new
nodes are attached preferentially to nodes that are already well
connected. The probability $\Pi(i)$ that a new node will be
connected to node $i$ is proportional to $k_i$, the degree of node
$i$.\\

\begin{equation}
\Pi(i) = {k_i\over \sum_j k_j} \label{eq:BA}
\end{equation}

Using the mean--field theory, Barab\'asi~{\sl et~al}
\cite{Barab99b} estimated that the BA model generates networks
with degree distribution of $P(k)\propto k^{-3}$. This model has
generated great interests in various research areas and has been
used as a starting point in the research of error and attack
tolerance of the Internet \cite{Albert00}.

\subsection{Generalized Linear Preference (GLP) model}

\begin{figure}[htb]
\begin{minipage}[t]{75mm}
\psfig{figure=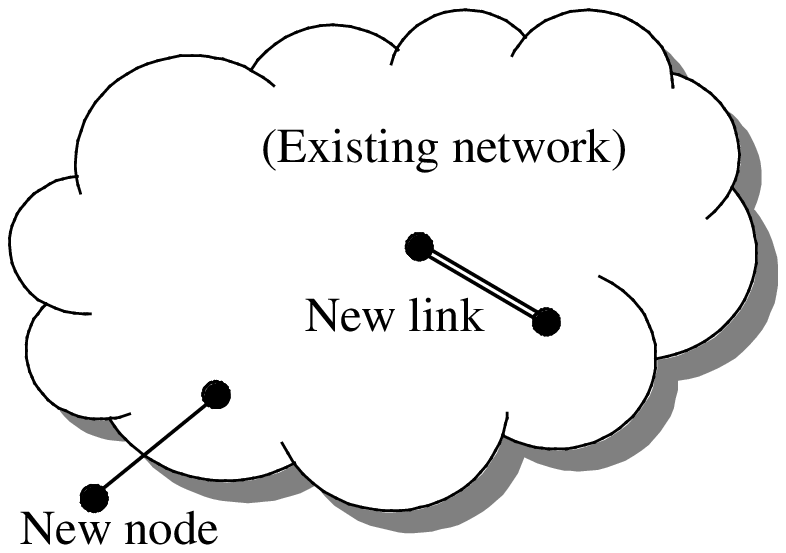,width=8cm}
\caption{The growth of
new nodes and new links are independent in the GLP model.}
\label{fig:GLP}
\end{minipage}
\hspace{\fill}
\begin{minipage}[t]{75mm}
\psfig{figure=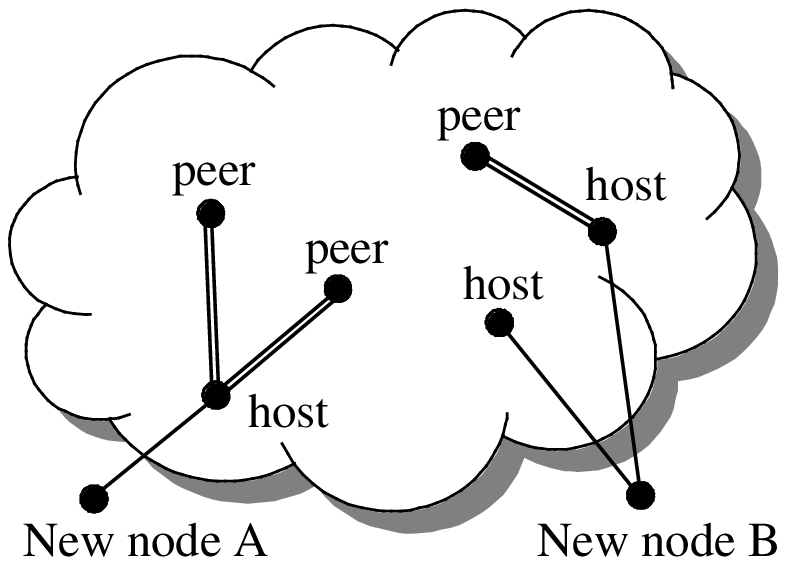,width=8cm}
\caption{The growth of
new nodes and new links are inter-dependant in the IG model.}
\label{fig:IG}
\end{minipage}
\end{figure}

The GLP model \cite{Bu02} was recently introduced. This model is a
modification of the BA model. It reflects the fact that the
evolution of the AS graph is mostly due to two operations, the
addition of new nodes and the addition of new links between
existing nodes. It starts with $m_0$ nodes connected through
$m_0-1$ links. As shown in Figure~\ref{fig:GLP}, at each
time--step, one of the following two operations is performed: 1)
with probability $\rho\in[0,1], m<m_0$ new links are added between
$m$ pairs of nodes chosen from existing nodes, and 2) with
probability $1-\rho$, one new node is added and connected to $m$
existing nodes. The GLP model uses the generalized linear
preference that the probability $\Pi(i)$ to choose node $i$ is
\begin{equation}
\Pi(i) = {k_i-\beta)\over \sum_j( k_j-\beta)} \label{eq:GLP}
\end{equation}
where the parameter $\beta\in(-\infty, 1)$ can be adjusted such
that nodes have a stronger preference of being connected to high
degree nodes than predicted by the linear preference of the BA
model (Equation~\ref{eq:BA}). This model matches the AS graph
(original--maps measured in Sept. 2000) in terms of the two
characteristic properties of small--world networks \cite{Watts98},
which are the characteristic path length and the clustering
coefficient.

\section{INTERACTIVE GROWTH (IG) MODEL}\label{section: IG model}

The Interactive Growth (IG) model also reflects the two main
operations that account for the evolution of the AS graph, the
addition of new nodes and the addition of new links. However the
growth of links and nodes are inter-dependant in the IG model. As
shown in Figure~\ref{fig:IG}, at each time--step, a new node is
connected to existing nodes (host nodes), and new links will
connect the host nodes to other existing nodes (peer nodes). The
IG model uses the same linear preference as the BA model
(Equation~\ref{eq:BA}) when choosing existing nodes to connect
with.

In the actual Internet, new nodes bring new traffic load to its
host nodes. This results in both the increase of traffic volume
and the change of traffic pattern around host nodes and triggers
the addition of new links connecting host nodes to peer nodes in
order to balance network traffic and optimize network performance.
We call the joint growth of new nodes and new links the
Interactive Growth (IG).

The joint growth of new nodes and new links has two significant
impacts, 1) rich nodes of the IG model are better inter-connected
to each other than those of the BA model; and 2) rich nodes of the
IG model have higher degrees than those of the BA model.

\begin{figure}[htb]
\begin{minipage}[t]{160mm}
\psfig{figure=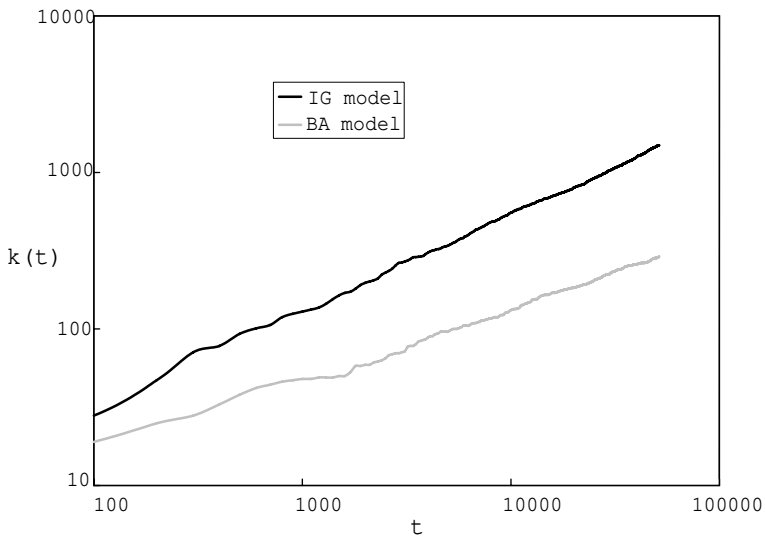,width=10cm}
\caption{Time--evolution
for node degree $k(t)$ against time $t$.} \label{fig:IG2}
\end{minipage}
\end{figure}

Figure~\ref{fig:IG2} shows that the time--evolution of node degree
in both the BA model and the IG model obeys a power--law
$k(t)\propto t^{\theta}$ . As predicted by Barab\'asi~{\sl et~al}
\cite{Barab99b}, $\theta$ of the BA model is $0.5$. Our
calculation shows that $\theta$ of the IG model is $0.6$. This
means that node degree in the IG model increases at a higher rate
than in the BA model. The reason is that during the interactive
growth of the IG model, host nodes not only connect to new nodes
but also acquire new links connecting to peer nodes.

\section{MODEL VALIDATION}

As shown in Table~\ref{table:1}, we generate 5 networks using the
above four models with the same number of nodes as the AS graph,
which in this paper is an extended--map measured on 26th May,
2001.

\begin{table}[htb]
\caption{Network properties} \label{table:1}
\renewcommand{\tabcolsep}{0.7pc} 
\renewcommand{\arraystretch}{1.2} 
\begin{tabular}{c r r r r r rrr}
\hline
~&              $N$     & $L$   & $k_{max}$ & $k_{average}$&$P(k=1)$&$P(k=2)$&$P(k=3)$\\
\hline
AS graph        & 11461 & 32730 & 2432  & 5.7 & 28.9\% & 40.3\% &11.6\% \\
IG model        & 11461 & 34363 & 842   & 6.0 & 26.0\% & 33.8\% &10.5\% \\
GLP(1)          & 11461 & 34363 & 517   & 6.0 & 68.4\% & 11.3\% & 5.1\%\\
GLP(2)          & 11461 & 34363 & 524   & 6.0 & 52.0\% & 16.3\% & 7.9\%\\
Inet--3.0       & 11461 & 24171 & 2010  & 4.2 & 40.0\% & 36.7\% & 8.2\%\\
BA model        & 11461 & 34363 & 329   & 6.0 & 0\%    & 0\%    & 40.0\%\\
\hline
\end{tabular}\\[2pt]
$N$ -- total number of nodes; $L$ -- total number of links;
$k_{average}$ -- maximum degree; $k_{average}$ -- average degree;
$P(k)$ -- degree distribution, percentage of nodes with degree
$k$.
\end{table}

The GLP model(1) is grown with parameters of $\rho=0.66$,
$m=1(m_0=10)$ and $\beta= 0.6447$, which is recommended by authors
of the model. The GLP model(2) is grown with the same parameters
except $\beta= 0$, which makes its generalized linear preference
equivalent to the linear preference of the BA model.

The IG model starts with a random graph of $m_0=10$ nodes and
$m_0$ links. In order to match the details of (low--range) degree
distribution of the AS graph, at each time--step, 1) with 40\%
probability, a new node (as node $A$ shown in Figure~\ref{fig:IG})
is connected to \emph{one} host node and the host node is
connected to \emph{two} peer nodes; and 2) with 60\% probability,
a new node (as node $B$ shown in Figure~\ref{fig:IG}) is connected
to \emph{two} host nodes and \emph{one} of the host nodes is
connected to \emph{one} peer node. Thus, $m=3$ new links are added
at every time--step.

\subsection{DEGREE DISTRIBUTION}
Degree distribution $P(k)$ is the percentage of nodes with degree
$k$. It is reported \cite{Chen02} that, as shown in
Figure~\ref{fig:Degree}, the degree distributions of the AS graph
deviates significantly from a strict power law. To take this into
consideration, we study the details of degree distribution by
examining the low--range degree distribution ($k\le 3$), the
high--range degree distribution (1000 richest nodes) and the
maximum degree, which is the largest number of links a node has.

\subsubsection{Low--range degree distribution}
The low--range degree distribution is important because nodes with
degree one and two account for more than 70\% of the total nodes
of the AS graph. Furthermore in the AS graph, the percentage of
nodes with degree one $P(1)$ is actually smaller than the
percentage of nodes with degree two $P(2)$.

\begin{figure}[htb]
\begin{minipage}[t]{75mm}
\psfig{figure=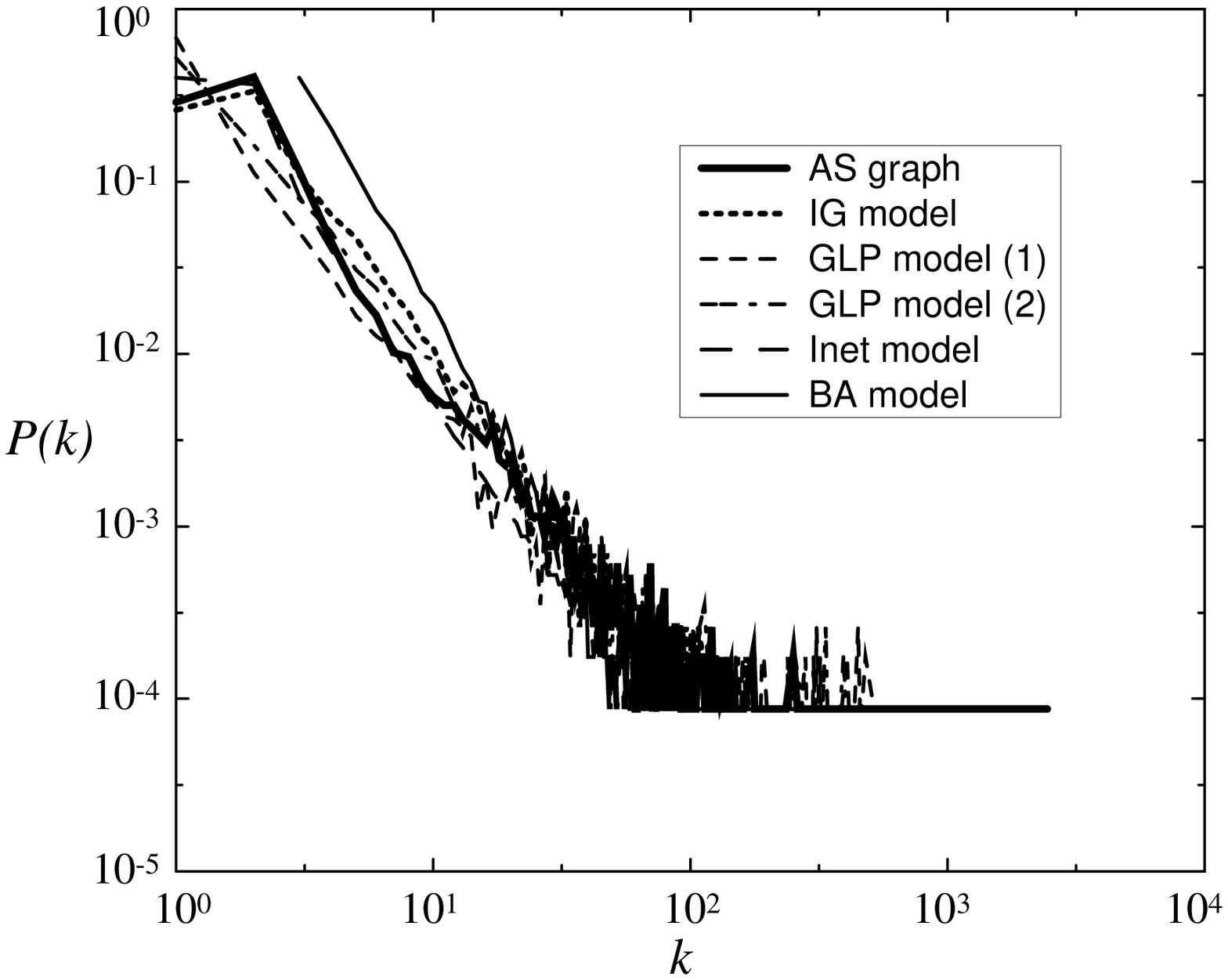,width=8cm}
\caption{Degree
distribution.} \label{fig:Degree}
\end{minipage}
\hspace{\fill}
\begin{minipage}[t]{75mm}
\psfig{figure=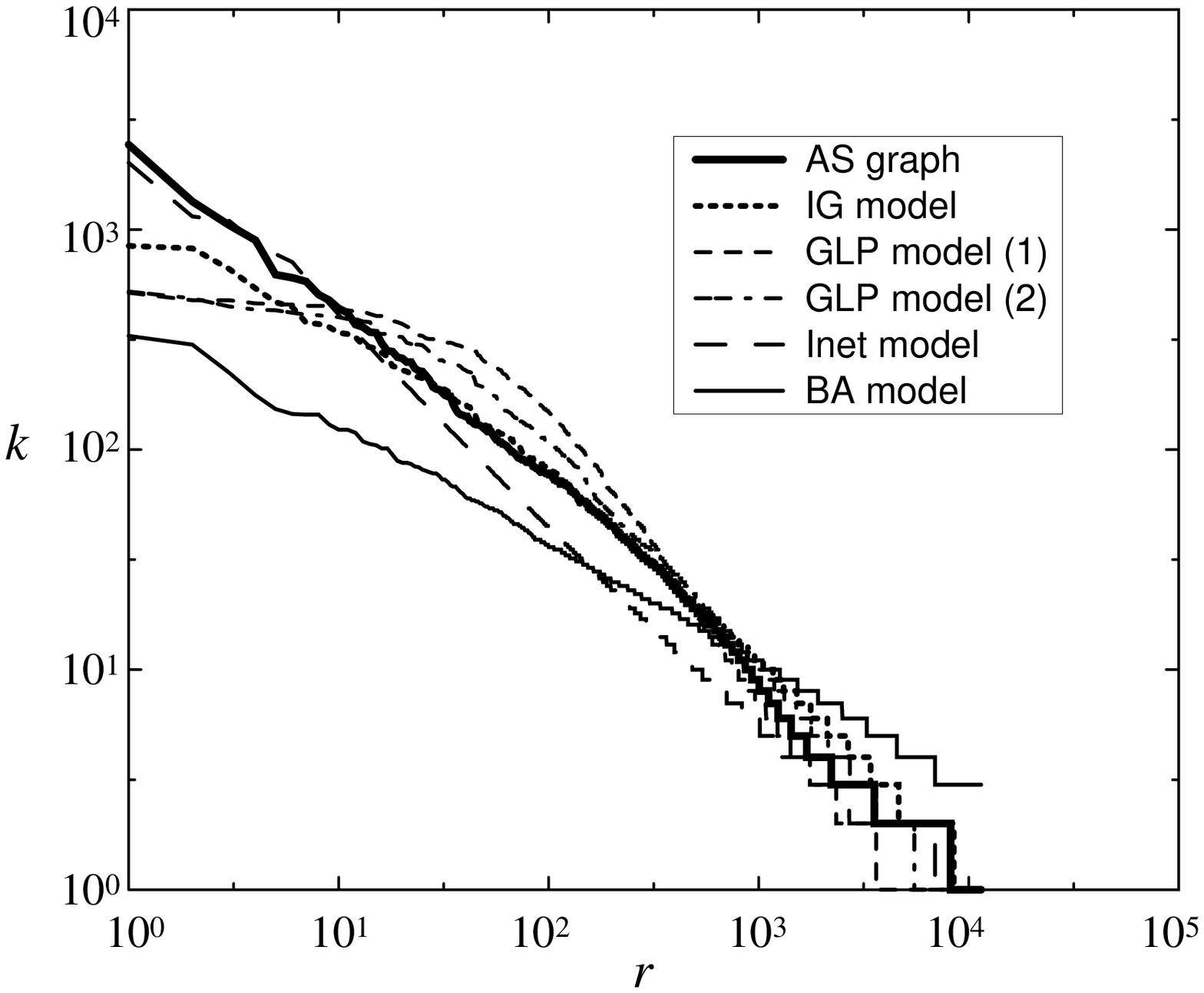,width=7.8cm}
\caption{Node degree $k$
against rank $r$.} \label{fig:Rank}
\end{minipage}
\end{figure}

Figure~\ref{fig:Degree} and Table~\ref{table:1} show that the IG
model and the Inet--3.0 model closely match the low--range degree
distribution of the AS graph. Only in the IG model, $P(1)$ is
smaller than $P(2)$. Whereas, other models do not resemble the
low--range degree distribution of the AS graph. For example,
$P(1)$ of the GLP model (1) is as high as 68.4\%.

\subsubsection{High--range degree distribution}

Figure~\ref{fig:Rank} is a plot of node degree $k$ against node
rank $r$, where $r$ is the rank of a node on a list sorted in a
decreasing order of node degree. Figure~\ref{fig:Rank} shows that
the high--range degree distributions $(r\le10^3)$  of the AS graph
are quite different from those of the two GLP models and the BA
model. The curve of the Inet-3.0 model significantly deviates from
that of the AS graph between $ r=10^1 $ and $ r=10^3 $. Apart from
the several richest nodes $ (r\le10^1)$, the IG model in general
well matches the high--range degree distribution of the AS graph.

\subsubsection{Maximum degree}
The AS graph feature a large value of maximum degree
($k_{max}=2432$). As shown in Table~\ref{table:1}, the GLP model
(1) with parameter $\beta= 0.6447$ and the GLP model(2) with
$\beta= 0$ (equivalent to linear preference) have similar maximum
degrees (517/524). This suggests that, although the generalized
linear preference of the GLP model (1) can make nodes have a
stronger preference of being connected to high degree nodes, it
does not effectively increase the maximum degree of the model.

The IG model uses the linear preference of the BA model. As we
mention in Section \ref{section: IG model} that due to the model's
interactive growth, the maximum degree of the IG model is
significantly higher than that of the BA model and the two GLP
models.

\subsection{Rich--club phenomenon}
The rich-club phenomenon is characterized by the rich-club
connectivity, which measures the interconnection between rich
nodes, and the node-node link distribution.

\subsubsection{Rich--club connectivity}
The maximum possible number of links between $n$ nodes is
$n(n-1)/2$. The rich-club connectivity $\phi(r)$ is defined as the
ratio of the actual number of links over the maximum possible
number of links between nodes with rank less than $r$, where $r$
is normalized by the total number of nodes.

\begin{figure}[htb]
\begin{minipage}[t]{75mm}
\psfig{figure=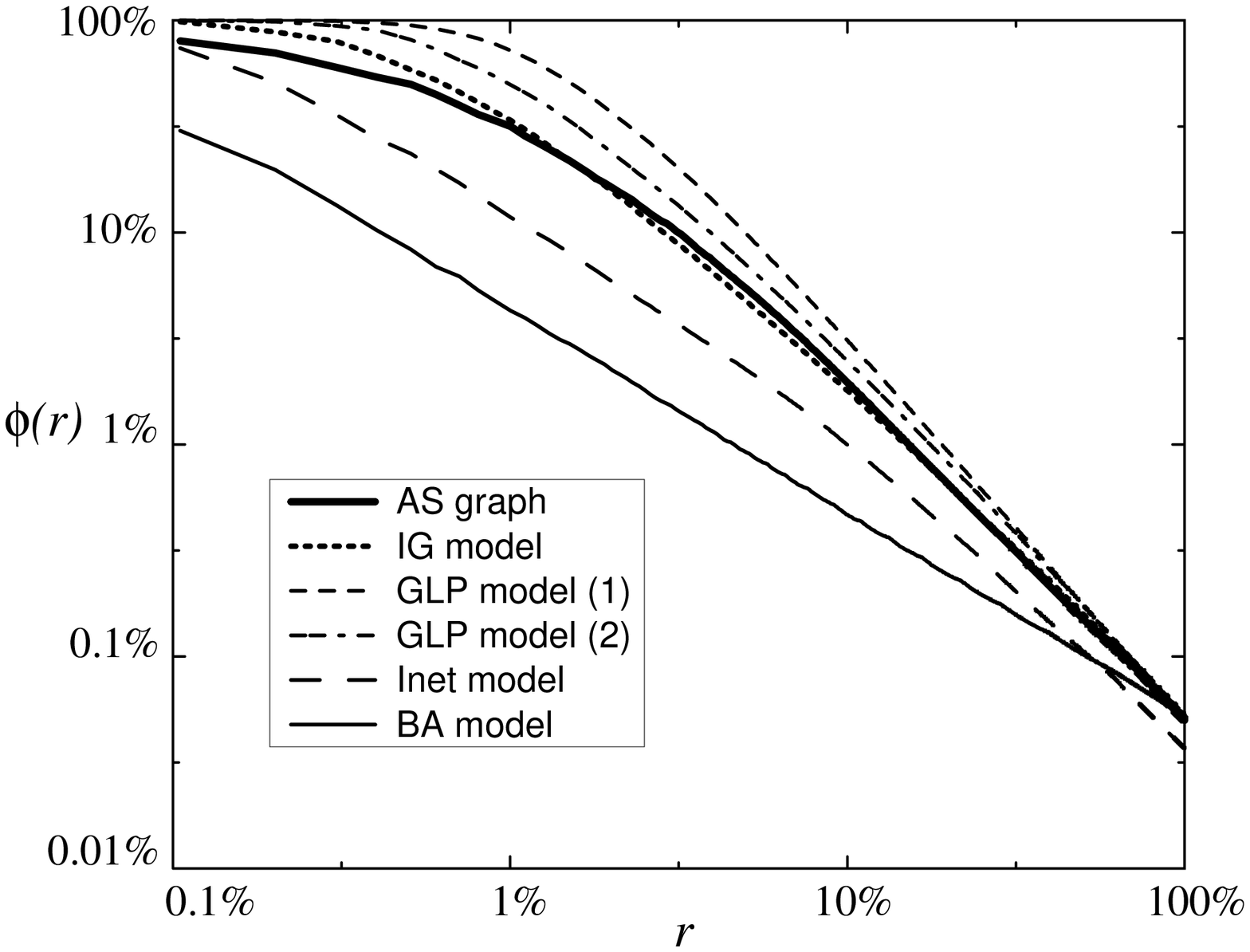,width=8cm}
\caption{Rich--club
connectivity $\phi(r)$ against node rank $r$.}
\label{fig:connectivity}
\end{minipage}
\hspace{\fill}
\begin{minipage}[t]{75mm}
\psfig{figure=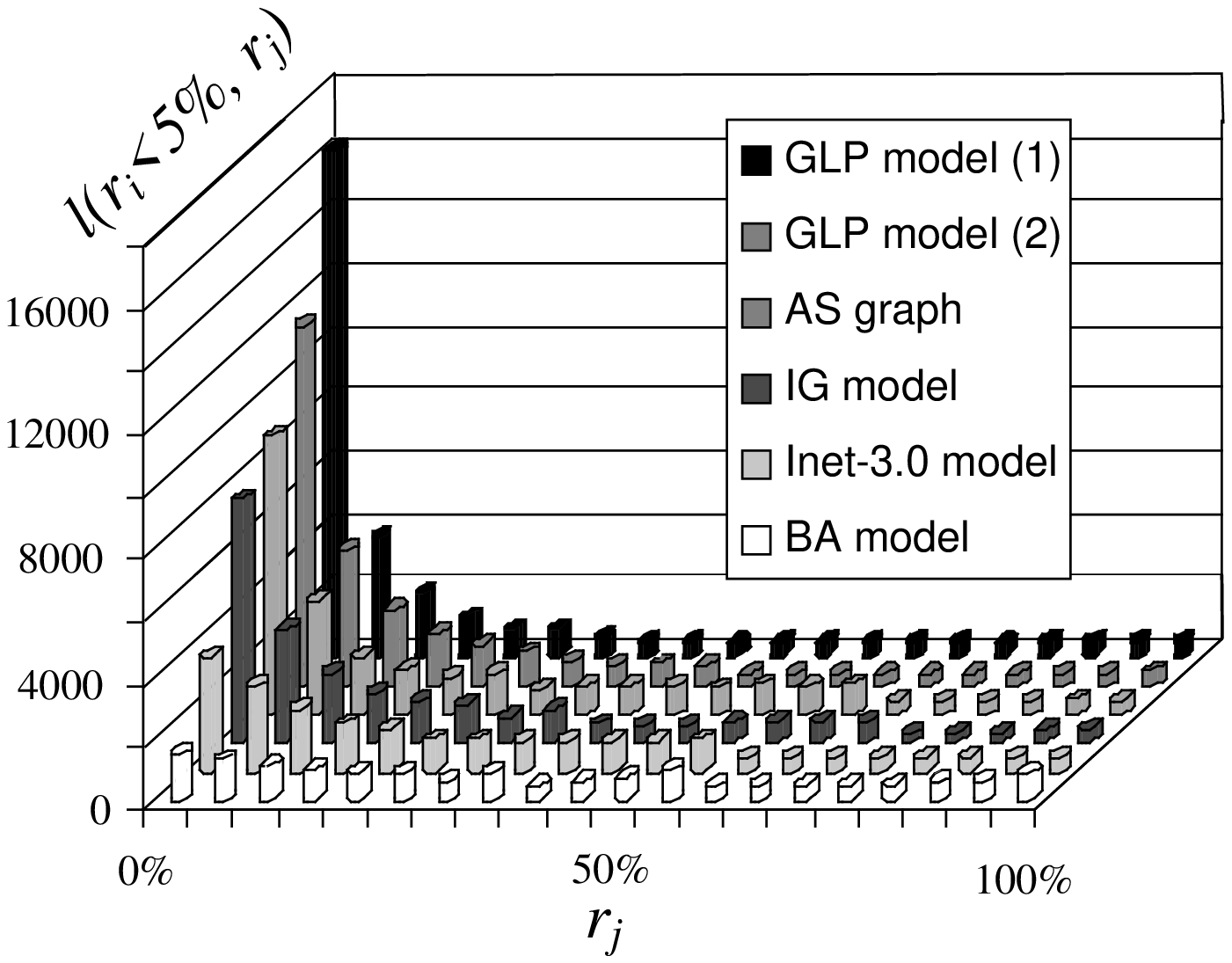,width=8cm}
\caption{Node-node link
distribution, number of link $l(r_i\le5\%, r_j)$ against node rank
$r_j$.} \label{fig:LinkDistribution}
\end{minipage}
\end{figure}

Figure~\ref{fig:connectivity} is a plot of the rich--club
connectivity $\phi(r)$ against node rank $r$ on a log--log scale.
The plot shows that only the IG model accurately matches the
rich--club connectivity of the AS graph. The rich--club
connectivity of the Inet--3.0 model and the BA model are
significantly lower than that of the AS graph. This means that
rich nodes in the two models are not as well connected to each
other as in the AS graph. On the other hand, the rich--club
connectivity of the two GLP models are significantly higher than
that of the AS graph. This means the rich nodes in these two
models are even more densely connected to each other than in the
AS graph. To be specific, the AS graph and the IG model have
$\phi(1\%)=32\%$, which indicates that the top 1\% richest nodes
have 32\% of the maximum possible number of links connecting
between them, comparing with $\phi(1\%)=18\%$ of the Inet--3.0
model, $\phi(1\%)=5\%$ of the BA model, $\phi(1\%)=72\%$ of the
GLP model(1) and $\phi(1\%)=50\%$ of the GLP model(2).

\subsubsection{Node-node link distribution}

We define $l(r_i , r_j)$ as the number of links connecting nodes
with node rank $r_i$ to nodes with $r_j$, where node ranks are
normalized by the total number of nodes and $r_i \le r_j$.

\begin{table}[htb]
\caption{Node--node link distribution} \label{table:2}
\renewcommand{\tabcolsep}{0.8pc} 
\renewcommand{\arraystretch}{1.2} 
\begin{tabular}{ccccccc}
\hline
~ & AS graph & IG & GLP(1) & GLP(2) & Inet & BA\\
\hline
$L$                         & 32730 & 34363 & 34363 & 34363 & 24171 & 34363 \\
$l(r_i\le5\%)$         & 29602 & 26422 & 32376 & 29073 & 22620 & 15687 \\
$l(r_i\le5\%, r_j\le5\%)$   & 8919 & 7806 & 16210 & 11540 & 3697 & 1511 \\
 \hline
\end{tabular}\\[2pt]
$L$ -- total number of links;\\
$l(r_i\le5\%)$ -- number of links connecting {\em with} the top
5\%
rich nodes;\\
$l(r_i\le5\%, r_j\le5\%)$ -- number of links connecting {\em
between} the top 5\% rich nodes.
\end{table}

Table~\ref{table:2} shows that the majority of links are the links
connecting with the top $5\%$ rich node, $l(r_i\le5\%)$.
Figure~\ref{fig:LinkDistribution} illustrates more details by
plotting $l(r_i\le5\%, r_j)$ against corresponding node rank
$r_j$, where $r_j$ is divided into $5\%$ bins.

The AS graph shows a rich--club phenomenon, in which rich nodes
are connected preferentially to the other rich nodes. The number
of links connecting between the top $5\%$ rich nodes
($l(r_i\le5\%, r_j\le5\%)=8919$) is significantly larger than the
number of links connecting these rich nodes to other lower degree
nodes.

The Inet--3.0 model does not show this phenomenon as strong as the
AS graph. The BA model does not show this phenomenon at all, in
stead, rich nodes are connected to nodes of all degrees with
similar probabilities. On the contrary, the two GLP models show
this phenomenon significantly stronger than the AS graph. In the
GLP model(1), $l(r_i\le5\%, r_j\le5\%)=16210$ is nearly twice of
that of the AS graph.

As seen from Figure~\ref{fig:connectivity}, Table~\ref{table:2}
and Figure~\ref{fig:LinkDistribution}, only the IG model
accurately reproduce the rich--club phenomenon of the AS graph.

\section{DISCUSSIONS AND CONCLUSIONS}

The Inet--3.0 model is not a dynamic model. Although it well
resembles the degree distribution, the model generates networks
with typically 26\% less links than the extended--AS graph.
Figure~\ref{fig:LinkDistribution} shows that the majority of the
missing links are connecting links between the rich nodes of the
AS graph. As a result, the Inet-3.0 model does not show the
rich-club phenomenon as strong as the AS graph.

The BA model generates a strict power--law degree distribution,
which is very different from that of the actual AS graph.
Moreover, it does not show the rich--club phenomenon of the AS
graph at all. This means that the network structure of the BA
model is fundamentally different from that of the AS graph.

The reason for this is that, according to the dynamics of the BA
model, all new links connect with new nodes. Due to the
preferential attachment, the probability for a new node to become
a rich node decreases as the network grows. This means that the
number of links between rich nodes rarely increases after the
initial growth. As a result, rich nodes are not well connected to
each other, and in the end, they are connected to nodes of all
degrees with similar probabilities. This result agrees with the
study of Chen~{\sl et al} \cite{Chen02} which showed that from the
available historical data, the AS graph does not evolves as  the
dynamics assumed in the BA model.

The GLP model does not reproduce the details of the degree
distribution of the AS graph. It is interesting to notice that the
rich--club phenomenon obtained from the GLP model is significantly
stronger than the AS graph.

Tangmunarunkit~{\sl et al} \cite{Tangm02} suggested that
degree-based network topology generators can match the degree
distribution and the hierarchy in real networks. Our results show
that the degree-based models can generate networks with
significantly different tier structures measured by the rich-club
phenomenon.

The IG model compares favorable with other Internet power-law
topology generators. It not only closely resembles the degree
distribution of the AS graph, but also accurately matches the
hierarchical structure of the AS graph. The IG model is a simple
and dynamic model. The topological properties of the networks
generated by the model are given by its joint growth of new nodes
and new links.

It is possible to include other parameters in the IG model, such
as bandwidth and delay. We expect the model to be used in
simulation-based research for the Internet traffic engineering. We
believe that the IG model is a good step towards modelling the
Internet topology.


\begin{thebibliography}{30}

\bibitem{Falou99}
M.~Faloutsos, P.~Faloutsos, and C.~Faloutsos, On Power--Law
Relationship of the Internet Topology. Proc. of ACM SIGCOMM 1999.

\bibitem{Subra02}
L.~Subramanian, S.~Agarwal, J.~Rexford and R.~H.~Katz,
Characterizing the Internet Hierarchy from Multiple Vantage
Points. Proc. of INFOCOM 2002.

\bibitem{Zhou03}
S.~Zhou and R.~J.~Mondragon, The missing links in the BGP--based
AS connectivity maps. Proc. of PAM2003.

\bibitem{Tangm02}
H.~Tangmunarunkit, R.~Govindan, S.~Jamin, S.~Shenker and
W.~Willinger, Network topology generators: Degree--based vs.
structural. Proc. of ACM SIGCOMM 2002.

\bibitem{Barab99}
A.~L.~Barab\'asi and R.~Albert, Emergence of Scaling in Random
Networks. Science, vol. 286, pp. 509--512, Oct. 1999.

\bibitem{Winick02} J.~Winick, and S.~Jamin, Inet--3.0: Internet Topology Generator,
Tech Report UM--CSE--TR--456--02. University of Michigan, 2002.

\bibitem{Bu02} T.~Bu and D.~Towsley, On Distinguishing between Internet Power Law
Topology Generators. Proc. of INFOCOM 2002.

\bibitem{Oregon}
University~of~Oregon, Route~Views~Project,
\emph{http://www.routeviews.org}.

\bibitem{Chen02}Q.~Chen, H.~Chang, R.~Govindan, S.~Jamin, S.~J.~Shenker
 and W.~Willinger, The Origin of Power Laws in Internet
Topologies (Revisited). Proc. of IEEE INFOCOM 2002.

\bibitem{Albert00} R.~Albert, H.~Jeong, and A.~L.~
Barab\'asi, Error and attack tolerance of complex networks.
Nature, vol. 406, pp. 378--381, July 2000.

\bibitem{Barab99b}
A.~L. Barab\'asi, R.~Albert, and H.~Jeong, Mean--field theory for
scale--free random networks. Physica A, vol. 272, pp. 173--187,
1999.

\bibitem{Watts98}
D.~J. Watts and S.~H.~Strogatz Collective dynamics of `small
world' networks, Nature, vol. 393, pp. 440--442, 1998.

\end{thebibliography}
\end{document}